\documentclass[aps,prd,floats,cite,preprint,tightenlines,nofootinbib]{revtex4}

\begin{document}

\preprint{\vbox{\hbox{HUTP-02/A033} 
    \hbox{MIT-CTP-3291} \hbox{hep-ph/0207243} }}
\vspace*{1cm}
\title{Little Higgses from an Antisymmetric Condensate}
\author{Ian Low $^a$, Witold Skiba $^b$, and David Smith $^b$}
\affiliation{\small $^a$Jefferson Physical Laboratory, Harvard University, Cambridge, MA 02138 \\
$^b$Center for Theoretical Physics, 
Massachusetts Institute of Technology, Cambridge, MA 02139 \vspace{1.5cm}}


\begin{abstract}
We construct an $SU(6)/Sp(6)$ non-linear sigma model in which the Higgses
arise as pseudo-Goldstone bosons. There are two Higgs doublets whose masses
have no one-loop quadratic sensitivity to the cutoff of the effective
theory, which can be at around 10 TeV. The Higgs potential is generated
by gauge and Yukawa interactions, and is distinctly different from that
of the minimal supersymmetric standard model. At the TeV scale, the new
bosonic degrees of freedom are a single neutral complex scalar and a
second copy of $SU(2)\times U(1)$ gauge bosons. Additional
vector-like pairs of colored fermions are also present.
\end{abstract}

\maketitle

\newpage

\section{Introduction}

There are many ways to embed the standard model in a more complete theory
which is valid at energies beyond a few TeV.  Those that remain weakly
coupled at the
TeV scale are preferred by precision electroweak data, and low-scale
supersymmetry is certainly
the best studied example in this class of theories.
Another potentially weakly coupled extension of the standard model, 
featuring the Higgs as a pseudo-Goldstone boson, has been
explored in far less detail. In order for the pseudo-Goldstone 
Higgs (PGH) to be weakly coupled in the TeV region, the effective
theory describing the PGH must have a cutoff somewhat larger than
a few TeV. Early attempts at constructing PGH
models~\cite{composite1,composite2} could achieve a high cutoff scale only
by fine tuning of parameters, because the Higgs mass suffered from a
quadratic sensitivity to the cutoff just as
in the standard model. Motivated by deconstruction~\cite{ACG1,HPW},
a PGH with a larger cutoff scale has been realized recently~\cite{ACG2}.
In the model of Ref.~\cite{ACG2}, the Higgs mass is quadratically
sensitive to the cutoff only at two loops. Using the set of
observations described in~\cite{ACG2,ACGW} it would be possible to
construct models where the sensitivity to the cutoff is suppressed
up to even higher orders in perturbation theory~\cite{ACG2}, but
it is not a high priority to do so since infrared-dominated contributions
to the Higgs mass prevent raising the cutoff scale arbitrarily high.
In any case, the cutoff can be pushed to around 10~TeV simply by
eliminating one-loop quadratic divergences, and we will not probe
beyond this scale anytime soon.

The interactions that contribute most to the quadratic divergence of the
Higgs mass are the top Yukawa coupling, the gauge interactions, and the Higgs
quartic coupling. How is it possible to eliminate the one-loop divergences
arising from these interactions? The key is to arrange for a large
approximate global symmetry which yields the desired pseudo-Goldstone
bosons after the symmetry is spontaneously broken.
Contained in this global symmetry is a product gauge group, such as
$(SU(2)\times U(1))^2$ or $SU(2)\times U(1) \times SU(3)^n$,
which is broken to the electroweak gauge group
when the global symmetry is spontaneously broken.
Each individual interaction, for instance a single $SU(2)$ group,
explicitly breaks only part of the global symmetry.
The remaining unbroken subgroup is sufficient to produce an exactly
massless Higgs. Therefore, the Higgs mass can arise only from an interplay
between two, or more, interactions and cannot become quadratically divergent
at one loop. Similar observations apply to fermion interactions.

The first models to incorporate this strategy for suppressing dominant
one-loop quadratic divergences~\cite{ACG1,ACGW,twosite} were based on
a QCD-like pattern of symmetry breaking (``moose models'').
The global symmetries of these models consist of multiple copies
of $SU(k)\times SU(k)$ spontaneously broken to $SU(k)$,
in analogy to chiral symmetry breaking in QCD with $k$ flavors.
Very recently, a model based on a simple global symmetry group
was constructed using the breaking pattern
$SU(5)\rightarrow SO(5)$~\cite{ACKN}. This model has very few
additional particles below the cutoff scale: a complex triplet of scalars,
an extra copy of $SU(2) \times U(1)$ gauge bosons, and a vector-like
pair of colored Weyl fermions. The model of Ref.~\cite{ACKN}
also has several other interesting features such as the generation of the
Higgs potential by the gauge interactions and the top Yukawa couplings.
These theories, in which the lightness of the Higgs is due to its
being a pseudo-Goldstone boson, have become known as ``little Higgs''
theories. 

Moose models based on a  QCD-like symmetry breaking pattern can be
analyzed in a systematic manner~\cite{GW} using notions of topology to predict
the low-energy spectrum and interactions. A similar understanding
of models with simple global symmetry groups is lacking. Since these
models offer interesting alternatives for TeV scale physics, it is
important to understand their general features. One way to approach
the problem is to study alternative models and check which properties
appear to be generic and which properties are model dependent.
In this article we study a model based on an $SU(6) \rightarrow Sp(6)$
symmetry breaking pattern.

Fermion condensation is one possible origin for this symmetry breaking.  
Condensed fermions that
transform under a complex representation of a strong gauge group could
yield a non-linear sigma model that is based on the coset $SU(k)\times
SU(k)/SU(k)$, where $k$ is the number of strongly interacting fermions.
A breaking pattern of this type is employed in the ``minimal moose''
model of Ref.~\cite{twosite}. When the condensed fermions transform under a
real representation of the strong gauge group, the condensate
gives rise to a non-linear sigma model based on $SU(k)/SO(k)$, which is
considered in Ref.~\cite{ACKN}.
The third possibility is that the strongly interacting
fermions transform under a pseudoreal representation, 
which produces the  $SU(k) \rightarrow Sp(k)$ breaking pattern
that interests us here\footnote{Of course, at the level of the
effective field theory, there is no need to restrict oneself to cosets 
that arise from fermion condensation.}.  
In this case the fermion condensate is antisymmetric in flavor.

However, there are a variety of possibilities for the underlying
physics of our model,  and the question of what dynamics generates 
the breaking of $SU(6)$ to $Sp(6)$ is not essential for our purposes. 
From the low-energy point of view, all available
information about the model is encoded in the nonlinear sigma-model
describing the coset space $SU(6)/Sp(6)$. If we made some assumptions
about the physics at the cutoff we might hope to compute, perhaps by using
lattice techniques, the coefficients of operators generated at
the cutoff. We will simply use naive dimensional
analysis\cite{Manohar:1983md} to estimate the 
magnitudes of the coefficients of any such operators.

In our model, the spontaneous breaking of $SU(6)$ to $Sp(6)$
generates $14=35-21$ (pseudo)Goldstone bosons. Four of these
become the longitudinal components of vector bosons, since
the condensate breaks the gauge group of the 
model, $(SU(2)\times U(1))^2$, to the electroweak $SU(2)_W\times U(1)_Y$.
Eight others  form two doublets of $SU(2)_W$ and remain light because
they are protected from one-loop quadratic divergences. The remaining two
form a complex neutral singlet. This singlet is not protected from
quadratic divergences and is heavier than the Higgs fields.

Therefore, at low energies our model is a two Higgs doublet model.
At energies of order TeV there is a set of massive gauge bosons
responsible for canceling one-loop divergences originating from the
gauge sector, along with the neutral singlet responsible for 
softening the divergences arising form the Higgs quartic coupling.
To eliminate the divergences associated with the top Yukawa we also
add extra colored fermions in vector-like pairs. Under the electroweak
symmetry these fermions transform as two doublets and two pairs of charged
singlets. We will explain in detail why this particular assignment is
selected.  As in the model of Ref.~\cite{ACKN}, the gauge and Yukawa
interactions are sufficient for generating a viable Higgs potential.

The outline of the rest of the paper is as follows:
in Section~\ref{sec:model}, we present the symmetry breaking pattern,
the embedding of the gauge group, and the effect of approximate symmetries
on the masses of the pseudo-Goldstone bosons. We also
describe the symmetry arguments which make keeping track of quadratic
divergences transparent. Section~\ref{sec:Yukawa} contains
a discussion of the top Yukawa coupling and the Higgs potential.
Section~\ref{sec:uv} considers the issue of vacuum stability in
the context of various possible UV completions.
Section~\ref{sec:pheno} contains a brief discussion of experimental
constraints and signatures. 

\section{The gauge sector}
\label{sec:model}

We consider a non-linear sigma model whose sigma field transforms under global
$SU(6)$ transformations $V$  according to  
\begin{equation}
\Sigma \rightarrow V \Sigma V^T.
\label{eq:sigmatrans}
\end{equation}
As stressed in the introduction, we will be interested in this model as a
low-energy effective field theory and will not specify the underlying physics
that provides its UV completion. We assume a condensate of the form
\begin{equation}
\langle \Sigma \rangle=
\left( 
\begin{array}{cc}
0 & -I \\
I & 0
\end{array}
\right),
\label{eq:sigmavev}
\end{equation}
where $I$ is the $3 \times 3$
unit matrix. The global $SU(6)$ symmetry is thus spontaneously broken
down to $Sp(6)$. 

We can divide the $SU(6)$ generators into
broken ($X_a$) and unbroken ($T_a$) ones. They satisfy
\begin{eqnarray}
X_a \langle \Sigma \rangle -\langle \Sigma \rangle X_a^T & = &  0,
\label{eq:broken}\\
T_a \langle \Sigma \rangle +\langle \Sigma \rangle T_a^T & = & 0.
\label{eq:unbroken}
\end{eqnarray}
The fluctuations of the sigma field around its vacuum expectation value
can then be parametrized by Goldstone boson fields $\Pi_a$ as in
\begin{equation}
\Sigma=e^{ i \Pi_a X_a/(2 f)} \langle \Sigma \rangle e^{ i \Pi_a X_a^T/(2 f)}
      =e^{i \Pi_a X_a/f} \langle \Sigma \rangle .
\label{eq:fluctuations}
\end{equation}
Using Eq.~(\ref{eq:broken}) we determine that the matrix of Goldstone
bosons has the form
\begin{equation}
\Pi_a X_a \sim
\left(
\begin{array}{cc}
A & B \\
B^\dagger & A^*
\end{array}
\right),
\end{equation}
where $A$ is a traceless and Hermitian $3\times 3$ matrix and $B$ is an antisymmetric $3\times 3$ matrix, giving
a total of 14 real fields.

We assume that $\langle \Sigma \rangle$ breaks an $(SU(2)\times U(1))^2$
gauge group 
to the electroweak $SU(2)_W \times U(1)_Y$. In our model the $SU(2)^2$ is a
subgroup of 
the global $SU(6)$ but the $U(1)^2$ is not, for reasons that will become
clear later. 
The full gauge group is contained in a global $U(6)$, and could also be
embedded 
in a larger group,
such as $SU(8)$, in a way consistent with our goals.
Let us label the two $SU(2)$ groups with a subscript $i=1,2$.
The generators of $SU(2)_i$ are 
\begin{equation}Q_1^a = {1\over 2}
\left(
\begin{array}{c}
{\begin{picture}(84,84)
\put(14,70){\makebox(0,0){\Large $\sigma^a$}}
\put(35,70){\makebox(0,0){0}}
\put(35,49){\makebox(0,0){0}}
\put(14,49){\makebox(0,0){0}}
\put(0,42){\line(1,0){84}}   
\put(42,0){\line(0,1){84}} 
\put(0,56){\line(1,0){42}} 
\put(28,42){\line(0,1){42}}
\put(63,63){\makebox(0,0){\LARGE 0}}
\put(63,21){\makebox(0,0){\LARGE 0}}
\put(21,21){\makebox(0,0){\LARGE 0}}
\end{picture}}
\end{array}
\right) 
\hspace{30pt} {\rm and} \hspace{30pt}
Q_2^a=-{1\over 2}\left(
\begin{array}{c}
{\begin{picture}(84,84)
\put(56,28){\makebox(0,0){\Large ${\sigma^a}^*$}}
\put(77,28){\makebox(0,0){0}}
\put(77,7){\makebox(0,0){0}}
\put(56,7){\makebox(0,0){0}}
\put(0,42){\line(1,0){84}}   
\put(42,0){\line(0,1){84}} 
\put(42,14){\line(1,0){42}} 
\put(70,0){\line(0,1){42}}
\put(63,63){\makebox(0,0){\LARGE 0}}
\put(21,63){\makebox(0,0){\LARGE 0}}
\put(21,21){\makebox(0,0){\LARGE 0}}
\end{picture}}
\end{array}
\right).
\label{eq:su2s}
\end{equation}
It is easy to check, using Eqs.~(\ref{eq:broken}) and (\ref{eq:unbroken}),
that the sum of $Q_1$ and $Q_2$ is unbroken, while the difference is broken.
The unbroken linear combination generates $SU(2)_W$.  The Goldstone bosons
which coincide with the broken gauged generators become the longitudinal
components  
of the massive gauge bosons of the broken $SU(2)$. 

The gauging of $(SU(2)\times U(1))^2$ explicitly breaks all of the
global $SU(6)$ symmetry, and the $\Pi_a$'s are not exact Goldstone bosons.
It is useful to decompose the pseudo-Goldstone states into 
representations of $SU(2)_W$, giving
\begin{equation}
\Pi_a X_a=
\left(
\begin{array}{c}
{\begin{picture}(120,120)
\put(0,60){\line(1,0){120}}   
\put(60,0){\line(0,1){120}} 
\put(0,20){\line(1,0){120}} 
\put(100,0){\line(0,1){120}}
\put(0,80){\line(1,0){120}} 
\put(40,0){\line(0,1){120}}
\put(50,100){\makebox(0,0){$\phi_1$}}
\put(110,100){\makebox(0,0){$\phi_2$}}
\put(20,70){\makebox(0,0){$\phi_1^\dagger$}}
\put(20,10){\makebox(0,0){$\phi_2^\dagger$}}
\put(110,40){\makebox(0,0){$\phi_1^*$}}
\put(80,10){\makebox(0,0){$\phi_1^T$}}
\put(80,70){\makebox(0,0){$-\phi_2^T$}}
\put(50,40){\makebox(0,0){$-\phi_2^*$}}
\put(20,100){\makebox(0,0){\LARGE $0$}}
\put(80,40){\makebox(0,0){\LARGE $0$}}
\put(50,70){\makebox(0,0){$0$}}
\put(110,10){\makebox(0,0){$0$}}
\put(70,110){\makebox(0,0){$0$}}
\put(90,90){\makebox(0,0){$0$}}
\put(110,70){\makebox(0,0){$0$}}
\put(10,50){\makebox(0,0){$0$}}
\put(30,30){\makebox(0,0){$0$}}
\put(50,10){\makebox(0,0){$0$}}
\put(90,110){\makebox(0,0){$s$}}
\put(70,90){\makebox(0,0){$-s$}}
\put(30,50){\makebox(0,0){$-s^*$}}
\put(10,30){\makebox(0,0){$s^*$}}
\end{picture}}
\end{array}
\right).
\label{eq:pions}
\end{equation}
Here, $\phi_1$ and $\phi_2$ are both $SU(2)_W$ doublets and $s$ is an
$SU(2)_W$ singlet.  We have set the eaten fields to zero,
and anticipated that the breaking of the gauged $U(1)^2$ to $U(1)_Y$
will eat the remaining diagonal piece of the Goldstone boson matrix
left over from the Higgsing of $SU(2)^2\rightarrow SU(2)_W$.  
Of the 14 real fields identified earlier, four are eaten, four each
are contained in $\phi_1$ and $\phi_2$, and two are contained in $s$.  

The fields $\phi_1$, $\phi_2$, and $s$ all acquire masses through their
gauge interactions, but the $\phi_i$'s
do not obtain one-loop mass contributions quadratic in the cutoff scale.
We will show that this is the case by  analyzing
the subgroup unbroken by each individual interaction,
and finding that the breaking of the exact symmetry produces massless
weak doublets. This type of reasoning~\cite{ACKN} will
be central to the rest of the paper,
and in particular will allow us to incorporate 
Yukawa couplings in the following section.

If either of the $SU(2)$
gauge couplings $g_1$ or $g_2$ vanish, the theory possesses an exact $SU(4)$
global symmetry. Suppose that $g_2$ is turned off, then $SU(4)$ acting on the
indices (3456) is an exact symmetry; if $g_1$ is turned off, an $SU(4)$
acting on (1236) is exact. Both of these $SU(4)$'s generate two weak
doublets under the breaking by $\langle \Sigma \rangle$.
This indicates that neither $\phi_1$ nor $\phi_2$ receives a quadratically
divergent mass squared at one loop. To generate a mass term for the
doublets both gauge interactions have to participate, so a quadratically
divergent contribution can only occur at two loops. On the other hand,
the $s$ scalar is not protected by the $SU(4)$ global symmetries and
should acquire a quadratically divergent mass squared at one loop.

The quadratically divergent gauge contribution to the Coleman-Weinberg
potential is 
\begin{equation}
{3\over 32 \pi^2}\Lambda^2 \, {\rm tr}\left[ M^2(\Sigma)\right], 
\end{equation}
where $\Lambda$ is the cutoff of the non-linear sigma model and
$M^2(\Sigma)$ is the gauge boson mass squared matrix in the presence of a
background value 
for $\Sigma$.  This matrix can be deduced from the kinetic term of the
sigma model Lagrangian  
\begin{equation}
-{f^2 \over 4}{\rm tr}[(D_\mu \Sigma)(D^\mu \Sigma)^*],
\label{eq:kinetic}
\end{equation}
where 
\begin{equation}
  D_\mu \Sigma=\partial_\mu \Sigma +ig_1{A_1}_\mu^a
    (Q_1^a \Sigma +\Sigma {Q_1^a}^T)+
    i g_2{A_2}_\mu^a (Q_2^a \Sigma +\Sigma {Q_2^a}^T).
\end{equation}
For simplicity, we include only the $SU(2)^2$ gauge interactions for the
time being. The presence of this quadratically divergent contribution
tells us that we must include in the effective Lagrangian a potential of
the form
\begin{equation}
  {2\over 3}c f^4 \left( g_1^2 {\rm tr}[(Q_1^a \Sigma)(Q_1^a \Sigma)^*]
  + g_2^2 {\rm tr}[(Q_2^a \Sigma)(Q_2^a \Sigma)^*] \right),
\label{eq:quaddiv}
\end{equation}
where $c$ is a coefficient of order one (the factor of $2/3$ is included
for later convenience).  
The minus sign in Eq.~(\ref{eq:kinetic}) is required by the antisymmetry
of $\Sigma$, and has the consequence that the quadratically divergent
gauge contribution to the Coleman-Weinberg potential is negative. 
Whether the corresponding operators included in the effective Lagrangian
come with the same or opposite sign depends on the details of the UV
completion of the effective theory.  

Expanding Eq.~(\ref{eq:quaddiv}) up to  terms quadratic  in $s$ and quartic in the doublets yields the potential
\begin{equation}
  c f^2 \left( g_1^2 \left| s+{i \over 2 f} {\tilde \phi_2}^\dagger 
         \phi_1 \right|^2+ g_2^2 \left| s-{i \over 2 f} 
         {\tilde \phi_2}^\dagger \phi_1 \right|^2 \right),
\label{eq:olp}
\end{equation}
where ${\tilde \phi_2}=i\sigma_2 \phi_2^*$.
If  these are the only contributions to
the potential with coefficients of this magnitude, the stability of the
vacuum of Eq.~(\ref{eq:sigmavev}) requires $c>0$, which we assume to be true.  
Only $s$ is massive at this level, 
and it can be integrated out to yield a low-energy quartic term for the light Higgses,
\begin{equation}
c  {g_1^2 g_2^2 \over g_1^2+g_2^2} \left|{\tilde \phi_2}^\dagger \phi_1 \right|^2.
\end{equation}
An order one quartic coupling is generated by integrating out the singlet,
just as it is generated by integrating out the triplet in the model
of Ref.~\cite{ACKN}.

For this quartic term to be suitable for stabilizing the Higgs condensates,
$\phi_1$ and $\phi_2$ must have opposite hypercharge, or else the quartic
term vanishes in the direction that preserves the $U(1)$ of 
electromagnetism.  Moreover, the two $U(1)$'s we gauge must not  induce
one-loop  quadratically divergent corrections to the Higgs masses.
We choose the $U(1)$'s generated by $Y_1={\rm Diag}(0,0,1,0,0,0)$
and $Y_2={\rm Diag}(0,0,0,0,0,-1)$, which are broken by
$\langle \Sigma \rangle$ to ordinary hypercharge $Y={\rm Diag}(0,0,1,0,0,-1)$.
This choice gives $\phi_1$ and $\phi_2$ opposite hypercharges as desired,
and makes the singlet $s$ neutral. Neither $Y_1$ nor $Y_2$ respect the
global $SU(4)$ approximate symmetries that protect the Higgs masses from
quadratically divergent contributions from $SU(2)$ gauge interactions.
However, they each respect an $SU(5)$ global symmetry, and these $SU(5)$'s
also prevent one-loop quadratic divergences. In fact, in the limit of
either $SU(5)$ being preserved, $\phi_1$, $\phi_2$, {\em and} $s$
become exact Goldstone bosons, so there are no quadratically divergent
contributions to the one-loop potential proportional to the $U(1)$
coupling constants.  

The form of the potential in Eq.~(\ref{eq:olp}) could have been deduced using
the global symmetry transformation properties of $\phi_1$, $\phi_2$,
and $s$~\cite{ACKN}. $SU(4)_{1}$ preserves global symmetries that act on these fields as
\begin{eqnarray}
  \phi_1 &\rightarrow& \phi_1 +  \epsilon_1 + \cdots, \nonumber \\
  \phi_2 &\rightarrow& \phi_2 +  \epsilon_2 + \cdots, \nonumber \\
  s &\rightarrow& s - {i \over 2 f}({\tilde \epsilon_2}^\dagger \phi_1
     +{\tilde \phi_2}^\dagger \epsilon_1)+\cdots,
\end{eqnarray}
while $SU(4)_{2}$ preserves global symmetries that act as 
\begin{eqnarray}
  \phi_1 &\rightarrow& \phi_1 +  \eta_1 + \cdots, \nonumber \\
  \phi_2 &\rightarrow& \phi_2 +  \eta_2 + \cdots, \nonumber \\
  s &\rightarrow& s + {i \over 2 f}({\tilde \eta_2}^\dagger \phi_1
    +{\tilde \phi_2}^\dagger \eta_1) +\cdots,
\end{eqnarray}
where $SU(4)_i$ is the symmetry preserved by $SU(2)_i$.
Given these transformation laws, the quadratically divergent contributions to 
the one loop potential are forced to be proportional to  Eq.~(\ref{eq:olp}).

\section{The top Yukawa coupling}
\label{sec:Yukawa}

The structure of our model guarantees that the Higgses are protected from  
one-loop quadratic divergences coming from gauge interactions and self
couplings.  The challenge remains to couple $\Sigma$ to fermions in a
way that gives rise to a large top Yukawa coupling without inducing
quadratic divergences from these additional interactions.
For this purpose we consider  the following fermion content: the standard
model fermions are taken to be singlets under $SU(2)_1$ and $Y_1$, and
have the usual assignments under $SU(2)_2$ and $Y_2$.  Thus the third
generation quarks have the charge assignments $b^c({\bf 1}_0, {\bf 1}_{1/3})$,
$t^c({\bf 1}_0, {\bf 1}_{-2/3})$, and $Q({\bf 1}_0, {\bf 2}_{1/6})$.
To the standard model fermions we add a colored pair of vector-like doublets
and two colored pairs of vector-like singlets: 
$Q'({\bf 2}_{1/6}, {\bf 1}_0)+{Q'}^c({\bf 2}_{-1/6},{\bf 1}_0)$,
$\psi_1({\bf 1}_{1/2},
{\bf 1}_{1/6})+ \psi_1^c({\bf 1}_{-1/2},{\bf 1}_{-1/6})$, and
$\psi_2({\bf 1}_{0},{\bf 1}_{-{1/3}})+ \psi_2^c({\bf 1}_{0},{\bf 1}_{1/3})$. 
We denote in parenthesis transformation properties under the gauge symmetries.

We couple $\Psi$ to $\Sigma$ through two terms, both of which preserve
a subgroup of the global $SU(6)$ that protects the Higgs masses. 
Consider the Lagrangian terms 
\begin{equation}
  \lambda_1 f \left({\begin{array}{cccc} {Q'}^T & 
   \psi_1 & (i\sigma_2 Q)^T & 0 \end{array}}\right) \Sigma^{*} 
    \left( \begin{array}{c} 0\\ 0\\  0 \\ t^c  \end{array} \right)
  +\lambda_2 f \left({\begin{array}{cccc} 0 & 0  & Q^T 
      & 0 \end{array}}\right) \Sigma 
     \left( \begin{array}{c} i \sigma_2 {Q'}^c \\ \psi_1^c \\ 0 
        \\ \psi_2^c  \end{array} \right)+ {\rm h.c.}
\label{eq:yuk}
\end{equation}
Neither of these terms induces one-loop quadratic divergent Higgs masses: 
the first term respects the same global $SU(5)$ as does $Y_2$ and the second
respects the global $SU(4)_2$.  These terms are gauge invariant 
given the charge assignments listed above.
We also introduce Dirac masses linking the vector-like fermion pairs:
\begin{equation}
  \lambda_3 f {Q'}^T i \sigma_2 {Q'}^c+\lambda_4 f \psi_1^c \psi_1 +
  \lambda_5 \psi_2^c \psi_2.
\label{eq:dirac}
\end{equation}

Expanding the two terms in Eq.~(\ref{eq:yuk}) gives
\begin{equation}
   -\lambda_1 [f t^c \psi_1 -i t^c(\phi_1^\dagger Q'
   +\tilde{\phi_2}^\dagger Q)]+\cdots +{\rm h.c.} \label{eq:linearyuk}
\end{equation}
and
\begin{equation}
  \lambda_2 [f Q^T i\sigma_2 {Q'}^c+
  i Q^T(\phi_1^* \psi_1^c+\phi_2^* \psi_2^c)]+\cdots +{\rm h.c.} 
\label{eq:42mass}
\end{equation}
where ${\tilde \phi_2}=i \sigma_2 \phi_2^*$.
In the limit of unbroken electroweak symmetry, the first terms in
these expansions combined with Dirac masses
of Eq.~(\ref{eq:dirac}) yield the massless linear combinations  
\begin{eqnarray}
  t^c_0 & = &{\lambda_1 \psi_1^c+\lambda_4 t^c \over
  \sqrt{\lambda_1^2+\lambda_4^2}}\\
  Q_0 & = &{\lambda_2 Q'-\lambda_3 Q\over \sqrt{\lambda_2^2+\lambda_3^2}}.
\end{eqnarray}
These light states
are the ordinary right-handed top quark and left-handed doublet.
The second term in Eq.~(\ref{eq:linearyuk}) contains the Yukawa couplings
for the top quark, 
\begin{equation}
  i\frac{\lambda_1}{\sqrt{(\lambda_1^2+\lambda_4^2)(\lambda_2^2+\lambda_3^2)}}
  t^c_0 \left[ \lambda_2 (\lambda_4-\lambda_3) \phi_1
               -\lambda_3 \lambda_4 \tilde{\phi_2}\right]^\dagger Q_0.
\end{equation}
Therefore, an order one top Yukawa coupling is easily obtained if the 
various $\lambda$'s are order one. It is important that a linear
combination of $\phi_1$ and $\tilde{\phi_2}$ appears  in the Yukawa coupling 
rather than $\phi_1$ or $\tilde{\phi_2}$ alone, which is why the  term
in Eq.~(\ref{eq:yuk}) proportional to $\lambda_2$ is necessary.  In its 
absence, the couplings of Eqs.~(\ref{eq:dirac}) and
(\ref{eq:linearyuk}) respect a Peccei-Quinn symmetry 
that would forbid the mass term $(\phi_1^\dagger \tilde{\phi_2} +{\rm h.c.})$. 
In this case, the Higgs potential either would 
preserve electroweak symmetry or would possess a runaway direction.   

Only $\lambda_2$, and not $\lambda_1$, contributes to the quadratically
divergent piece of the Coleman-Weinberg potential, giving
\begin{equation}
  {3 \over 8 \pi^2}\Lambda^2 \left|s-{i \over 2 f} 
    {\tilde \phi_2}^\dagger \phi_1 \right|^2,
\end{equation}
Since $\lambda_2$ preserves $SU(4)_2$, its contribution is proportional to the
second term of Eq.~(\ref{eq:olp}). The presence of this additional 
quadratic divergence requires that the potential of Eq.~(\ref{eq:olp})
is modified to
\begin{equation}
c  g_1^2 f^2 \left|s+{i \over 2 f} {\tilde \phi_2}^\dagger \phi_1 \right|^2
+(c g_2^2+c' \lambda_2^2) f^2 \left|s-{i \over 2 f} 
{\tilde \phi_2}^\dagger \phi_1 \right|^2,
\end{equation} 
where $c'$ is another order one coefficient.  The mass squared of $s$ is
positive provided that $c' \lambda_2^2 + c (g_1^2+g_2^2)>0$.  In this case,
integrating out $s$ gives the low energy quartic coupling
$\lambda|{\tilde \phi_2}^\dagger \phi_1|^2$, with
\begin{equation}
\lambda=c { g_1^2 [g_2^2+(c'/c)\lambda_2^2]  \over g_1^2+ [g_2^2+(c'/c) \lambda_2^2]}.
\label{eq:quarticcoupling}
\end{equation}
Note that $\lambda>0$  requires $c>0$.

The quadratic terms in the one-loop Higgs potential are logarithmically
divergent and take the form\footnote{Because we have assumed for simplicity
that the coefficients in Eqs.~(\ref{eq:yuk}) and (\ref{eq:dirac}) are real,
$b^2$ will be real, but phase redefinitions  of the Higgs doublets could
have absorbed the phase of $b^2$ were it complex.} 
\begin{equation}
  m_1^2|\phi_1|^2+m_2^2 |\phi_2|^2 +b^2(\phi_1^\dagger 
  {\tilde \phi_2}+{\rm h.c.}).
\label{eq:quadratic}
\end{equation}
The quartic term in the Higgs potential features flat directions that
are stabilized only if both $m_1^2$ and $m_2^2$ are positive.  
In this case the potential is minimized for vacuum expectation values
that may be chosen to have the form
\begin{equation}
  \phi_1= \left(\begin{array}{c}0 \\ v_1 \end{array}\right) 
  \hspace{30pt}{\rm and}\hspace{30pt} \phi_2 = \left( \begin{array}{c} v_2 \\
  0 \end{array}\right),
\label{higgsvev}
\end{equation}
where $v_1$ and $v_2$ are both real, with the same sign if $b^2$ is
positive and opposite sign if $b^2$ is negative. Electroweak symmetry
breaking will occur if and only if the determinant of the Higgs mass
matrix is negative,
\begin{equation}
  m_1^2 m_2^2 - b^4 < 0.
\label{determinant}
\end{equation}

Fermion loops induce logarithmically divergent  contributions to 
$m_1^2$, $m_2^2$ and $b^2$.  Assuming the heavy fermion masses,
$M_{t'}=f\sqrt{\lambda_1^2+\lambda_4^2}$,  
$M_{Q'}=f\sqrt{\lambda_2^2+\lambda_3^2}$, and $M_{\psi_2}=f\lambda_5$, 
to be the same order of magnitude and, to an excellent approximation,
taking all the logarithms to a common value, $\log(\Lambda/M)$, 
these contributions are
\begin{eqnarray}
  {m_1^2}_f & = & {3 f^2 \over 8\pi^2}(\lambda_1^2 -\lambda_2^2)
      (\lambda_3^2 -\lambda_4^2) \log{\Lambda^2 \over M^2} \label{eq:m1}\\
  {m_2^2}_f & = &{3 f^2 \over 8\pi^2} 
      (\lambda_1^2 \lambda_2^2+\lambda_2^2 \lambda_5^2-
         \lambda_2^2 \lambda_3^2 -\lambda_1^2 \lambda_4^2) 
       \log{\Lambda^2 \over M^2}\label{eq:m2}\\
   b^2_f & = & {3 f^2 \over 8\pi^2}\lambda_1^2 \lambda_2 
       (\lambda_3 - \lambda_4) \log{\Lambda^2 \over M^2} .
\end{eqnarray}
Gauge and scalar loops contribute only to $m_1^2$ and  $m_2^2$, with
\begin{eqnarray}
  {m_1^2}_g={m_2^2}_g & = & {3 \over 64 \pi^2} 
     \left(3 g^2 M_g^2  \log{\Lambda^2 \over M_g^2}+
     {g'}^2 {{M_{g'}}^2 \log{\Lambda^2 \over {M_{g'}}^2}} \right),\\
  {m_1^2}_s={m_2^2}_s & = & \frac{\lambda}{16\pi^2} M_s^2 
     \log{\frac{\Lambda^2}{M_s^2}},
\end{eqnarray}
where $M_g=f\sqrt{(g_1^2+g_2^2)/2}$, $M_{g'}=f\sqrt{({g'}_1^2+{g'}_2^2)/2}$,
and $M_s=f\sqrt{c(g_1^2+g_2^2)+c'\lambda_2^2}$ are the masses of the
heavy $SU(2)$ gauge boson, the heavy $U(1)$ gauge boson, and $s$,
respectively.  The crucial ingredient in the above formulae is the
logarithmically divergent contribution to $b^2$ coming from fermion loops.
Because all the entries of the Higgs mass matrix have similar logarithmic
enhancement, there is a broad range of parameters for which 
Eq.~(\ref{determinant}) is satisfied and electroweak symmetry is broken.

There may be other terms in the effective Lagrangian, involving $\Sigma$
alone, that contribute to the Higgs potential. In this case, these extra
terms may be used to break the Peccei-Quinn symmetry, for example through
the gauge invariant operator
\begin{equation}
  a f^4\epsilon_{ij}\Sigma_{3i} \Sigma_{jx} \Sigma_{x3}^*+{\rm h.c.}=
    -(a f^2 {\tilde \phi}_2^\dagger \phi_1+{\rm h.c.})+\cdots,
\end{equation}
where $i,j=1,2$ and $x=4,5$, and $a$ is an order one coefficient.
It is natural for $a$ to be small ($\ll 16\pi^2$) because this term breaks
all of the global $SU(4)$'s and $SU(5)$'s identified earlier. If the
Peccei-Quinn symmetry is broken in this way, simpler possibilities arise
for the fermion sector.  For instance, one could remove $\psi_2$ and
$\psi_2^c$ from the previous model and do away with the term in
Eq.~(\ref{eq:yuk}) proportional to $\lambda_2$.  In this case the additional
fermionic matter content beyond the standard model is one vector-like pair
of singlets and one vector-like pair of doublets.

\section{Vacuum Stability and UV completions}
\label{sec:uv}

In this section we would like to address the issue of the stability
of the vacuum we have chosen, Eq.~(\ref{eq:sigmavev}), and possible UV
completions of the non-linear sigma model. As mentioned in the introduction,
one way to UV complete the effective theory is by using strong gauge dynamics
at the cutoff scale. The desired symmetry breaking pattern could arise from
a condensate of six Weyl fermions transforming under a pseudoreal
representation of a strong gauge group, for example the fundamental
representation of $Sp(N)$. If the strong group confines and breaks
chiral symmetry then the global $SU(6)$ symmetry associated with the
six flavors is spontaneously broken down to the $Sp(6)$ subgroup.

If we were to use strong dynamics to UV complete our model, we could
assign two of the six flavors to transform as a doublet of $SU(2)_1$,
another two flavors as a doublet of $SU(2)_2$. The remaining two flavors
would be $SU(2)_i$ singlets and carry charges $(1,0)$ and $(0,-1)$
under $U(1)_1\times U(1)_2$, respectively.  This assignment gives
the same charges for the nonlinear sigma model as described
in Sect.~\ref{sec:model}. The fermions charged under the $U(1)$
symmetries induce mixed anomalies with the strong group that can be
canceled by the Green-Schwarz mechanism. Alternatively, one could
enlarge the theory, for example to one based the $SU(8)/Sp(8)$ coset space,
to accommodate traceless $U(1)$ generators.

An important question is whether the vacuum we have chosen,
Eq.~(\ref{eq:sigmavev}),  is stable in a model with strong dynamics.
For the time being, let us concentrate on the operators induced by the
gauge interactions Eq.~(\ref{eq:quaddiv}). A generalization of the standard
argument\cite{alignment},  drawing on the observation that
$\pi^\pm-\pi^0$ mass difference is positive in QCD, would imply that
the coefficient $c$ is negative for the $SU(6)/Sp(6)$ model. 
In QCD-like theories, the true vacuum is aligned such that the sum
of the gauge boson masses squared is minimized, which selects not the
vacuum of Eq.~(\ref{eq:sigmavev}), but instead
$\langle \Sigma \rangle = \sigma_2^{12} \otimes \sigma_2^{36}
  \otimes \sigma_2^{45}$, which preserves both $SU(2)$'s. The superscripts
on the $\sigma_2$ matrices indicate the indices the $SU(6)$ indices. 
However, the dynamics of the $Sp(N)$ gauge theories could be
very different from that of the QCD-like theories, in which case a true
determination  of the sign of $c$ would require a lattice calculation. 
If the sign of $c$ is indeed negative, there are two options for
obtaining a viable UV completion. One possibility is to consider different
dynamics responsible for breaking the  global symmetry. It is likely,
for instance, that either  sign of $c$ is possible in a supersymmetric
UV completion.  Another possibility is that other interactions overwhelm
the gauge interactions and yield a stable vacuum. 

For concreteness, we discuss the second possibility in detail. For example,
it is possible for the fermion contributions to stabilize the potential
by introducing another vector-like pair of quarks
$Q''({\bf 1}_0,{\bf 2}_{1/6})+{Q''}^c({\bf 1}_0,{\bf 2}_{-1/6})$
and adding to Eq.~(\ref{eq:yuk}) the coupling
\begin{equation}
  \lambda_2^\prime f \left({\begin{array}{cccc} {Q'}^T & 0  & 0
  & 0 \end{array}}\right) \Sigma
  \left( \begin{array}{c} 
      0 \\ \psi_1^c \\ i \sigma_2 {Q''}^c \\ \psi_2^c
  \end{array} \right)
  + {\rm h.c.} \quad .
\end{equation} 
This term preserves an $SU(4)_1$ global symmetry,
whereas the $\lambda_2$ term in Eq.~(\ref{eq:yuk}) preserves a different
$SU(4)_2$ symmetry.
We also include in the Lagrangian all possible
gauge-invariant mass terms involving the quarks.
Then a linear combination of $Q$, $Q'$, and $Q''$ will
be massless before electroweak symmetry breaking, and the
top Yukawa coupling comes from the $\lambda_1$ term in Eq.~(\ref{eq:yuk})
as before.
Now the potential generated at the cutoff has the form
\begin{equation}
  (c  g_1^2+c'{\lambda_2^\prime}^2) f^2 \left|s+{i \over 2 f}
   {\tilde \phi_2}^\dagger \phi_1 \right|^2
  +(c g_2^2+c' \lambda_2^2) f^2 \left|s-{i \over 2 f}
   {\tilde \phi_2}^\dagger \phi_1 \right|^2.
\end{equation}
To achieve the electroweak symmetry breaking we need {\em both}
$c g_1^2+c'{\lambda_2^\prime}^2$ and $c g_2^2+c' \lambda_2^2$ to be 
positive. The stability of the vacuum follows.

Alternatively, the vacuum might be stabilized by operators involving $\Sigma$
alone.  We call these ``plaquette operators'' in analogy with Ref.~\cite{ACG2},
even though our operators have no apparent lattice interpretation.
For example, suppose that in the effective Lagrangian we include
\begin{equation} 
  a_1 f^4 \Sigma_{ij} \Sigma^*_{ji}+a_2 f^4 \Sigma_{xy} \Sigma^*_{yx},
\end{equation}  
with $i,j$ summed over $1,2$ and $x,y$ summed over $4,5$.
These operators yield a contribution to the potential
\begin{equation}
  2 f^2 a_1 \left|s+{i \over 2 f} {\tilde \phi_2}^\dagger \phi_1 \right|^2 
  +2 f^2 a_2 \left|s-{i \over 2 f} 
{\tilde \phi_2}^\dagger \phi_1 \right|^2.
\end{equation}
If $a_1$ and $a_2$ are larger than $c$ times the gauge couplings squared,
these terms can stabilize the vacuum and achieve the electroweak symmetry
breaking at the same time.

Note that the stability is accomplished by operators that either
contain two fermions and one power of the $\Sigma$ field, or 
two powers of $\Sigma$. Both can be generated by four-fermion
interactions and could easily come from an analog of
extended technicolor.

\section{Phenomenology}
\label{sec:pheno}

In addition to the matter and gauge fields of the standard model, our model
contains two light Higgs doublets. The other heavy states, with masses
below $\Lambda \approx 10$ TeV, are an $SU(2)_W\times U(1)_Y$-neutral
complex scalar, a copy of $SU(2)\times U(1)$ gauge bosons, a vector-like
pair of $SU(2)_W$-doublet colored fermions, and two vector-like pairs
of $SU(2)_W$-singlet colored fermions, all of which have masses around
a few TeV. The ordering of the spectrum of TeV-scale particles is
parameter dependent. Our consideration of these heavy states' effects
on the precision electroweak measurements parallels the analogous discussion
in~\cite{ACKN}. The number of electroweak doublet fermions which get mass
through Yukawa couplings is tightly constrained by the $S$
parameter\cite{Holdom:1990tc,Peskin:1990zt,Golden:1990ig}.
Fortunately, the additional fermions in our model are all vector-like,
which was shown in Refs.~\cite{Maekawa:1994yd,Maekawa:1995ha,Lavoura:1992np}
to have small contributions to both the $S$ and $T$ parameters. The effects
of these heavy, weakly interacting particles decouple from the low
energy physics and are suppressed by factors of $M_W^2/M_{\rm heavy}^2$.
Thus, their impact on the precision electroweak measurements should be
similar to 2-loop standard model corrections and smaller than the
current experimental errors. The scalar $s$ in our model is
an $SU(2)_W$ singlet and integrating it out at the tree level does not
induce custodial symmetry violating operators, as would be the case
were it a triplet scalar~\cite{ACKN}. Therefore, the contribution to
the $T$ (or $\rho$) parameter is minimal. Finally, the heavy
$SU(2) \times U(1)$ gauge bosons couple directly to standard model fermions,
and induce tree-level contributions to muon decay, for instance.
This places a lower bound on $M_g$ of roughly $3$ TeV~\cite{Marciano:1999ih}
if the heavy gauge bosons couple to  light fermions with the same strength
as the light gauge bosons, but if $g_1 \neq g_2$ this need not be the case
and the bound can be weakened.

We are required to introduce a vector-like pair of $SU(2)_W$ doublet
quarks to cancel the quadratically divergent contributions to the Higgs
masses coming from the top Yukawa coupling, so we predict that at the
TeV scale both charge 2/3 quarks and charge 1/3 quarks will be present.  
These can decay to ordinary top and bottom quarks by emitting $Z$, $W$,
or Higgs particles.  The $s$ scalar mixes with the Higgs doublets with
mixing angle approximately $v/f$, and decays predominantly into pairs
of Higgs bosons. It is unfortunately unlikely to be detected at the
LHC because it is neutral under $SU(2)_W\times U(1)_Y$, but more work
is needed to study possible production and decay channels of the singlet
scalar.

The two light Higgs doublets yield two CP-even neutral
scalars $h^0$, $H^0$, one CP-odd neutral scalar $A^0$, and two charged
scalars $H^+$ and $H^-$, with masses
\begin{eqnarray}
 M_{A^0}^2 &=& \frac{2 b^2}{\sin(2\beta)},  \nonumber \\
 M_{H^\pm}^2 &=& M_{A^0}^2 - \frac{\lambda v^2}2, \nonumber \\
 M_{h^0}^2,\, M_{H^0}^2 &=& \frac12 \left( M_{A^0}^2 \pm
    \sqrt{M_{A^0}^4 - 4 M_{H^\pm}^2 (M_{A^0}^2 - M_{H^\pm}^2)
\sin^2(2\beta)}\right),
\label{higgsmass}
\end{eqnarray}
where $v^2\sim 250$ GeV, and the ratio of vacuum expectation values,
$\tan \beta \equiv v_2/v_1$, turns out to be equal to $\sqrt{m_1^2/m_2^2}$.
It is interesting to note that the CP-odd scalar $A^0$ is heavier than
the charged scalars $H^\pm$.  This is contrary to the MSSM, a difference
that arises from the fact that the quartic potential of our model has a
different form than that of the MSSM. The lighter CP-even Higgs mass
$M_{h^0}$ is maximized when $\sin^2(2\beta)=1$ ($m_1^2=m_2^2$), in which
case the CP-even scalar masses are at tree level equal to $M_{H^\pm}^2$
and $M_{A^0}^2 - M_{H^\pm}^2$.

The light fermions must be coupled to the two Higgs doublets in such a way
that tree level flavor-changing neutral currents (FCNCs) are
avoided~\cite{Glashow:1976nt}, which can be accomplished either by
coupling all the fermions to a single Higgs doublet, or by coupling
up-type and down-type quarks to separate Higgs doublets. For instance,
the light fermion masses might come from
\begin{equation}
  y^{u}_{ij} f \left({\begin{array}{cccc} 0 & 0 & (i\sigma_2 Q_i)^T 
    & 0 \end{array}}\right) \Sigma^{*}    \left( \begin{array}{c} 0\\ 
    0\\  0 \\ u^c_j  \end{array} \right)+y^{d}_{ij} 
     f \left({\begin{array}{cccc} 0 & 0 & Q_i^T & 0 \end{array}}\right) 
  \Sigma  \left( \begin{array}{c} 0\\ 0\\  0 \\ d^c_j  \end{array} \right),
\end{equation}
in which case both up quarks and down quarks couple to $\phi_2$.
These couplings induce one-loop quadratic divergences in the Higgs potential,
but they are harmless because of the smallness of the light quark Yukawa
couplings. The only tree-level FCNCs induced by Higgs exchange involve the
top sector and so are not problematic.

Going beyond the low-energy effective theory, there could also be new,
potentially dangerous sources of FCNCs in some UV completions, for instance
if the fermion couplings to $\Sigma$ arise from four-fermi interactions as
in extended technicolor~\cite{Chivukula:2002ww}.  Possible ways of
suppressing these FCNCs are discussed in Ref.~\cite{Chivukula:2002ww,ACKN}.

\acknowledgments 
We are grateful to Nima Arkani-Hamed and Andy Cohen for numerous stimulating
discussions. We also thank Ami Katz and Nick Toumbas for helpful conversations.
I.L. is supported by the National Science Foundation under grant
number PHY-9802709, and  W.S. and D.S. are supported by the U.S. Department
of Energy under grant DE-FC02-94ER40818. W.S. and D.S. acknowledge the
hospitality of the Aspen Center for Physics where this work was completed.

\end{document}